# Quantum suppression of superconductivity in ultrathin nanowires

A. Bezryadin, C. N. Lau, and M. Tinkham

*Department of Physics, Harvard University, Cambridge, MA 02138*

**It is of fundamental importance to establish whether there is a limit to how thin a superconducting wire can be, while retaining its superconducting character, and, if so, what sets this limit. This issue may also be of practical importance in defining the limit to miniaturization of superconducting electronic circuits. It is well established that at high temperatures the resistance of linear superconductors is caused by excitations called thermally activated phase slips (TAPS)[1,2,3,4]. Quantum tunneling of phase slips is another possibility which is still being debated[5,6,7,8]. There is a theoretical prediction that such quantum phase slips (QPS) can destroy superconductivity in very narrow wires[8]. Here we report measurements on ultrathin ($\lesssim 10\,\text{nm}$) nanowires produced by coating carbon nanotubes with a superconducting MoGe alloy. We find that nanowires can be superconducting or insulating depending on their normal state resistance $R_N$ compared to $R_q = h/(2e)^2$ -- the quantum resistance for Cooper pairs. If $R_N < R_q$ the tunneling of QPS is prohibited due to strong damping, and so the wires stay superconducting. The insulating state, observed if $R_N > R_q$, is explained in terms of proliferation of QPS and corresponding localization of Cooper pairs.**

The phenomenon of superconductivity depends on the coherence of the phase ($j$) of the superconducting order parameter. For small systems, such as Josephson junctions (JJ) or ultrathin wires, $j$ is a quantum variable which may or may not have a definite value, corresponding to superconducting and insulating states respectively: The system becomes superconducting when its wave function becomes localized in the $j$-space. Properties of a JJ can be understood from an analogy with a quantum particle in a periodic potential[9], whose wave function is a delocalized Bloch wave. Since the Josephson energy is a periodic function of $\varphi$, the JJ should also be delocalized in the $\varphi$-space, and thus *insulating*, for arbitrary strength of the Josephson coupling. However, this is not always true since the JJ is a *macroscopic* quantum system[10] which interacts with environment. This interaction[11], when linear, can be described in terms of the classical friction coefficient ($\eta$). The friction can reduce the energy bandwidth to zero and localize the particle[12,13,14]. Such quantum localization transition is called dissipative phase transition (DPT)[15]. It occurs at some critical value of $\eta$ *independent* of the strength of the periodic potential. In the case of JJ[15], the dissipation is controlled by the normal conductance ($G_{NJJ}$), and the DPT appears as an insulator-superconductor (SI) transition at $G_{NJJ} = (2e)^2/h$. The physics of superconducting wires is more complicated since $\varphi$ can vary along the wire. The nature of SI transitions in nanowires has recently been analyzed theoretically by many authors[7,8,16], yet there is no consensus at present.

In this paper we investigate experimentally the possibility of a DPT in ultrathin wires. If it exists, it could be the main mechanism which controls superconductivity in nanowires. By analogy with JJ, we may expect that the DPT should be controlled by the wire's normal-state conductance ($G_N \equiv 1/R_N$) and should not depend explicitly on its diameter, which determines the energy barrier for phase slips. We have measured several nanowires of diameters $\lesssim 10$ nm, ranging in length from $L \approx 95$ nm to $L \approx 185$ nm, much longer than the coherence length $\xi \approx 8$ nm. It is found that the wires are superconducting only if $R_N$ is lower than the quantum resistance for Cooper pairs ($R_q = h/(2e)^2 \approx 6.5$ k$\Omega$), and insulating otherwise. This agrees with the DPT interpretation: At weak dissipation the tunneling of phase slips is dominant in ultrathin wires and destroys superconductivity. Nevertheless the superconductivity is recovered at $R_N < R_q$, when the dissipation, proportional to $G_N$, is strong enough to suppress the QPS tunneling.

Quantum effects are measurable only in ultrathin nanowires of size ~ 10 nm [8] which is below the resolution limit of electron beam lithography. We have developed a powerful new technique which allows fabrication of uniform nanowires considerably thinner than 10 nm (Fig.1c). This is achieved by sputtering a superconducting alloy of amorphous $Mo_{79}Ge_{21}$ over a free-standing carbon nanotube or bundle of tubes which is laid down over a narrow and deep slit[17] etched in the substrate (Fig.1a). The wire width is determined by the width of the underlying bundle, which serves as a template for metal deposition. Graybeal and Beasley[18] discovered that sputtered MoGe films are amorphous, have a sharp superconducting transition, and show no signs of granularity down to ~1 nm film thickness. Our nanowires are five times thicker, so they are expected to be very homogeneous. Indeed, even the narrowest wires of width $W \approx 5.5$ nm are continuous (see Fig.1c) with the surface roughness of ~1 nm.

The sample resistance was determined from the slope of the current-voltage (I-V) curves measured at a frequency of 0.48 Hz by current biasing the leads I+ and I- (Fig.1a). A low bias amplitude (4 nA) ensured the linearity of the I-V curves. The voltage was measured with a low-noise battery-operated amplifier PAR 113. Examples of resistance versus temperature curves are shown in Fig.2a. The bottom curve shows the superconducting transition of the electrodes. The top curve gives the resistance $R(T)$ of the nanowire in series with a section of the electrodes. Below $\approx 5.5$ K the electrodes are resistanceless but the wire is not, and $R(T)$ exactly equals the wire resistance. The resistance measured immediately below the film transition is taken to be the normal state resistance $R_N$ of the wire[19] (Fig.2a).

The homogeneity (absence of a granular structure) of the wires is established by measuring their normal state bulk resistivity. We find that the resistivity is the same for all wires and agrees well with the standard values for a-$Mo_{79}Ge_{21}$. Fig.2b shows a plot of the unit length normal conductance $G_0 \equiv L/R_N$ versus the wire width $W$. *All* points in Fig.2b, including the insulating samples, can be fit with a straight line. The slope gives the resistivity as $r = d/(dG_0/dW) \approx (1.8 \pm 0.4)$ $m\Omega \cdot m$. This is in excellent agreement with the values $r_f \approx 1.7$ $m\Omega \cdot m$ and $r_b \approx 1.65$ $m\Omega \cdot m$ measured on 5 nm thick a-$Mo_{79}Ge_{21}$ films[18] and on bulk samples respectively. In the above argument we assume that the film thickness is constant across the wire and equals the film thickness $d = 5$ nm. Some deviations from this model geometry are possible since the film on the supposedly circular nanotubes can have tapering edges. On the other hand, the deviations should be weak since the sputtering is far from being unidirectional. Also, the edge roughness is much smaller than the coherence length, so the superconducting effects should not be influenced. The following measurements also confirm that the wires are uniform over their length: (i) Residual resistance ratio for all wires is $RRR \geq 0.85$, similar to the standard macroscopic value $\approx 0.95$. (ii) A single-electron-tunneling gate effect has *not* been seen (doped Si substrate served as a gate). (iii) The critical current was ~1 µA (for s3) which is similar to the value expected for a uniform wire. (iv) The transition at the critical current is very sharp and shows no steps.

The results of transport measurements are summarized in Fig.3a. Two qualitatively different types of behavior are found: The samples s1, s2, s3, ss1, and ss2 appear to be superconducting since their resistance decreases ~exponentially with cooling. The wires i1, i2, and i3 (thick curves) do not show any signatures of a superconductivity (the resistance drop at $\approx 5.5$ K comes from the electrodes) and their resistance stays almost constant with cooling. Therefore these samples may be normal or even insulating. In fact, they should be considered insulating because: (i) their resistance increases at low enough temperatures (Fig.3b) and (ii) their differential resistance ($R_d \equiv dV/dI$) have a pronounced maximum[15] (Fig.3c, bottom curve) at zero bias current ($I$), and therefore $dR_d/dI < 0$ at small $I$. The superconducting samples showed positive derivatives $dR/dT > 0$ and $dR_d/dI > 0$ (see Fig.3c, top curve) in all cases, including very low temperatures: one sample has been tested down to 50 mK.

Fig.3a shows that the type of the $R(T)$ dependence is controlled by the normal resistance $R_N$ of the wire (see also Fig.2a). This behavior can be explained assuming that the wires undergo a superconductor-insulator (SI) quantum phase transition at $R_N = R_q$. This conclusion is supported by the following facts: (i) All samples with $R_N < R_q$ are superconducting while all samples with $R_N > R_q$ are insulating. (ii) The insulating vs. superconducting properties become stronger when the difference $|R_N - R_q|$ increases (except for the samples with two wires). For example, the temperature of the upturn in Fig. 3b increases with $R_N$. (iii) The difference between the superconducting and insulating samples becomes more pronounced at lower temperatures, indicating that the transition is driven by quantum (not thermal) fluctuations. Note that the observed dichotomy (Fig. 3a) resembles the SI transition in amorphous thin films (2D limit)[20] which occurs when their *resistance per square* $R_{\square} = r/d$ reaches $R_q \approx 6.5$ k$\Omega$. Despite apparent similarities, we think that our transition is different since our 5 nm thick MoGe film (which forms the leads *and* the wire) has a much lower resistance per square $R_{\square} \approx 350$ $\Omega << R_q$.

Two dashed curves (Fig. 3a) correspond to samples with two parallel wires. Accordingly they have the lowest normal state resistance. Nevertheless their low temperature tails are similar to single-wire samples with about twice higher $R_N$. This shows that the dissipative events (presumably the phase slips) are localized inside each wire and do not depend on the presence of another wire ~ 0.5 $m$m apart.

The important role of the normal state conductance $G_N \equiv 1/R_N$, evident from our measurements, suggests that the observed SI transition is a DPT. The amount of dissipation is controlled by the normal conductance $G_N$ because each phase slips generates a finite voltage pulse ($V_p$) on the wire. The dissipated power is simply $V_p^2 G_N$, i.e. proportional to the normal conductance. Each phase slip transforms the kinetic energy of Cooper pairs into thermal energy and causes heating. At zero temperature and zero bias current one has to consider virtual phase slips. This ensures that no real dissipation takes place when the system is in its ground state. In this case the damping, which is still proportional to the normal conductance, is also virtual.

As we already discussed, a quantum system becomes superconducting when the damping is strong enough to suppress tunneling in the $j$-space and to localize the phase. In contrast to the JJ case (which is geometrically a 0D system), the phase fluctuations in a nanowire (which is 1D) can vary in length from $x$ up to the length of the wire ($L$). The "friction" acting on a QPS depends on the length scale over which the phase change occurs: The longer the QPS the weaker the dissipation since the conductance is proportional to $1/L$. The tunneling of larger QPS can therefore be dominant, and the mean field theory, which considers only phase slips of size $x$ (which are energetically the most favorable) may not be sufficient to analyze the DPT. To accurately describe the QPS of all sizes one has to use the renormalization group (RG) theory of Zaikin et al.[8] and Demler et al. Qualitatively speaking, the QPS can be suppressed at all scales, up to $L$, only if the normal conductance of the whole wire exceeds the conductance quantum. Therefore, independently of the wire length, the DPT should take place at $G_N = (2e)^2/h$. Our measurements on long wires ($L \sim 20x$) appear to be in good agreement with the above conclusions, confirming that we observe a DPT. Although it may seem surprising that DPT in nanowires occurs at the same resistance ($R_N = R_q$) as in Josephson junctions, this could be a natural consequence of the critical point divergence of the length scale of fluctuations. When the spatial dimensions of the quantum fluctuations become larger than the length of the wire, the system (nanowire) becomes effectively zero-dimensional and therefore equivalent to JJ. This shows a similarity between quantum phase transitions observed in mesoscopic systems and thermodynamic phase transitions which occur in bulk systems.

The negative derivatives $dR/dT < 0$ and $dR_d/dI < 0$, measured on the insulating wires, can also be understood in terms of QPS. The insulating state is delocalized in the $j$-space and localized in the conjugate, charge space, meaning localization of the Cooper pairs. External perturbations such as thermal fluctuations or a nonzero bias voltage weaken the charge-localization effect, leading to the observed reduction in the differential resistance. The shape of the $dR_d/dI$ vs $I$ dependence measured for the insulating samples (Fig.3c) is very similar to measurements[15] on JJs which undergo a DPT. This similarity gives an additional confirmation that the transition which we find near $G_N = (2e)^2/h$ is a DPT.

In conclusion, we presented the first observation of a dissipative transition in superconducting nanowires which occurs near $G_N = (2e)^2/h$. The results indicate that at the transition the QPS length scale becomes equal to the wire length $L \gg x$. In the future we plan to investigate the DPT in longer wires and compare it to the RG theory of Zaikin et al.[8], which predicts that the wire diameter becomes important when the length exceeds $\sim hc_{MS}/k_BT \sim 10$ µm ($c_{MS}$ is the velocity of the Mooij-Schön mode[21]).

Correspondence and requests for materials should be addressed to A. B. (e-mail: alexey@rsj.harvard.edu)

## Acknowledgement

We would like to acknowledge very informative discussions with E. Demler, Y. Oreg and D. S. Fisher about their new QPS theory. Also we thank M. Bockrath, B. I. Halperin, L. Levitov, J.E. Mooij, C. van der Wal, R. M. Westervelt and A. D. Zaikin for discussions and S. Shepard for help with fabrication. This research was supported in part by the NSF and ONR.

**Figure captions**

**Figure 1** Fabrication and imaging of a nanowire. **a,** Schematic of the sample. Si substrate (black) is covered with a 0.5 µm layer of $SiO_2$ and a 50 nm thin SiN film. A 100 nm wide slit is patterned in the SiN film using electron beam lithography and reactive ion etching (RIE). The $SiO_2$ layer under the slit is removed using HF to form an undercut (white area). To make a nanowire we first deposit nanotubes (red) and then sputter a 5 nm thick $a\text{-}Mo_{79}Ge_{21}$ film and a 1.5 nm Ge protective layer. Initially the metal film covers the entire substrate including free-standing carbon nanotubes or bundles crossing the slit. Next we find a suitable single nanowire (using SEM) and pattern the electrodes (blue) using optical lithography and RIE. The MoGe film is interrupted by the slit and forms two electrodes connected electrically through a single nanowire. **b,** Scanning electron micrograph (SEM) of a sample. The image shows a free-standing nanowire which connects two MoGe electrodes. This is one of the thinnest wires found under the SEM. Its apparent width, including blurring in the SEM image, is $W \cong 10$ nm. The scale bar (white) is 200 nm. **c,** One of the thinnest wires found under a higher resolution transmission electron microscope (TEM). The wire width is $W_{TEM} \approx 5.5 \pm 1$ nm. The scale bar (black) is 100 nm.

**Figure 2** Normal state properties of the wire. **a,** Definition of the normal state wire resistance ($R_N$). The bottom curve is the resistance of the 5 nm thick MoGe film (the leads) vs temperature, measured on the contacts $V_f$ and V+ (see Fig.1a). The top curve is taken on V+ and V- contacts and shows the resistance $R(T)$ of a nanowire connected in series with a section of the leads. Below 5.5 K (vertical line) the leads are superconducting and the top curves gives the resistance of the nanowire. The normal state resistance $R_N$ (horizontal line) is measured immediately below the film transition where the wire is still normal. **b,** Unit length wire conductance $G_0 \equiv L/R_N$ versus the wire width $W$. The width is measured by SEM. $L$ is the effective length of the wire. Each data point represents a different sample. The data are shown for superconducting (circles)

and insulating samples (squares). The straight line is $G_0 = C(W - W_0)$ with $C = 2.8 \cdot 10^{-3}$ S and $W_0 = 7.9$ nm. Note that $W$ systematically overestimates the actual width of the conducting core of the wire due to SEM smearing (compare Fig.1b and Fig.1c) and due to the presence of the Ge protective film on top of each wire. This explains why the linear fit does not extrapolate to zero.

**Figure 3** Transport properties of superconducting and insulating nanowires. **a,** Resistance versus temperature curves for eight different samples. The superconducting transition of the leads takes place at $T_c \approx 5.5$ K. Sample ss1 has a different $T_c \approx 4.3$ K, presumably due to a different substrate treatment. Samples ss1 (dashed curve) and ss2 (short-dash curve) contain two parallel wires. All other samples have only a single wire. The samples i1, i2, i3, s1, s2, s3, ss1, ss2 have the following parameters: Apparent widths (measured under SEM) are $W =$ 11, 11.4, 13.2, 18, 21, 16.2, 13.3 and 15 nm. The wires in each of pair ss1 and ss2 have about the same width. The normal state resistances are $R_N = 22.6$, 14.79, 10.29, 6.42, 4.53, 5.66, 3.09, and 3.2 kOhm. The effective lengths $L(\text{nm}) =$ 185, 135, 130, 168, 165, 146, 57, and 72. For the pairs of parallel wires the quoted length is calculated as $1/L = 1/L_1 + 1/L_2$. The lengths of the ss1 wires are $L_1 = 96$ nm and $L_2 = 139$ nm. For the ss2 pair the lengths are: $L_1 \approx L_2 \approx 143$ nm. **b,** Magnification of the $R(T)$ curves i1, i2 and i3. The curves are vertically displaced for clarity. All three samples show $dR/dT < 0$ at low enough temperatures. The upturn takes place at $T =$ 3.2, 2.8, and 1.6 K for the samples i1, i2, and i3 respectively. **c,** A plot of normalized differential resistance ($R_d \equiv dV/dI$, $R_{d0} \equiv dV/dI|_{I=0}$) versus the bias current ($I$) for two samples: i1 ($T = 1.2$ K) and s1 ($T = 4.2$ K).

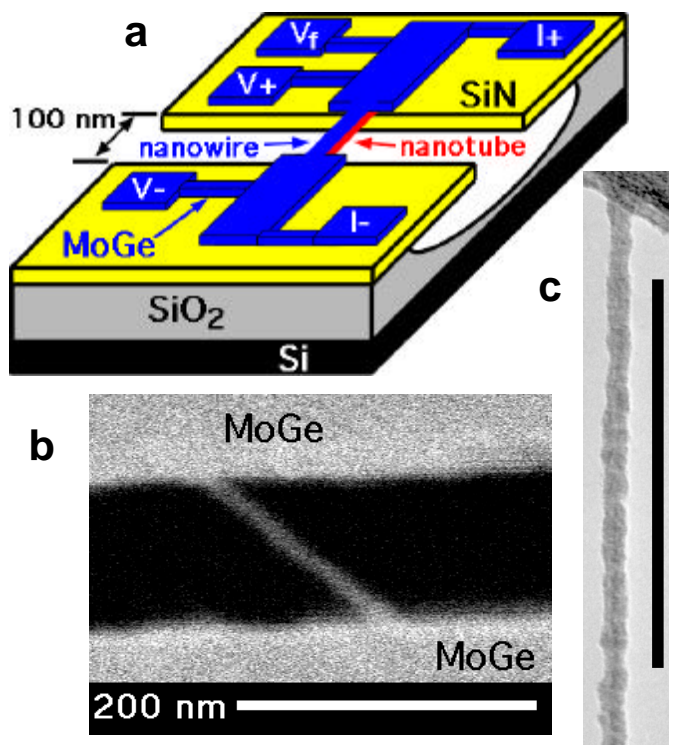

Fig.1

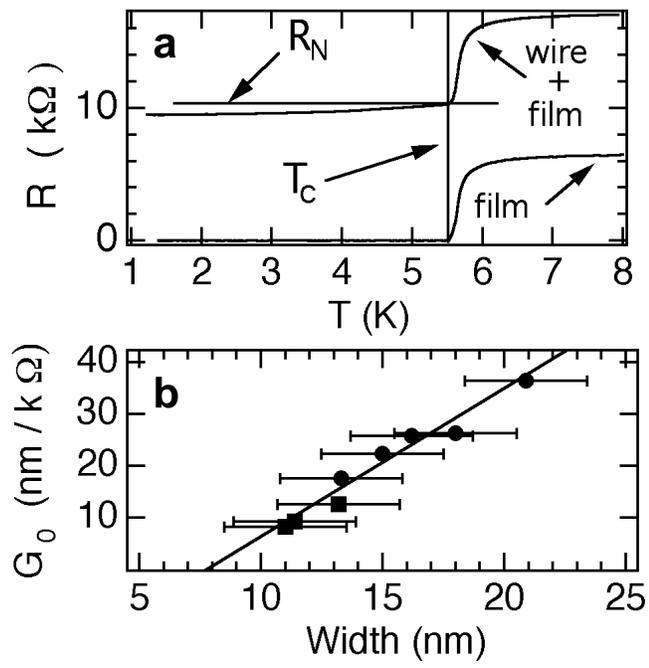

Fig.2

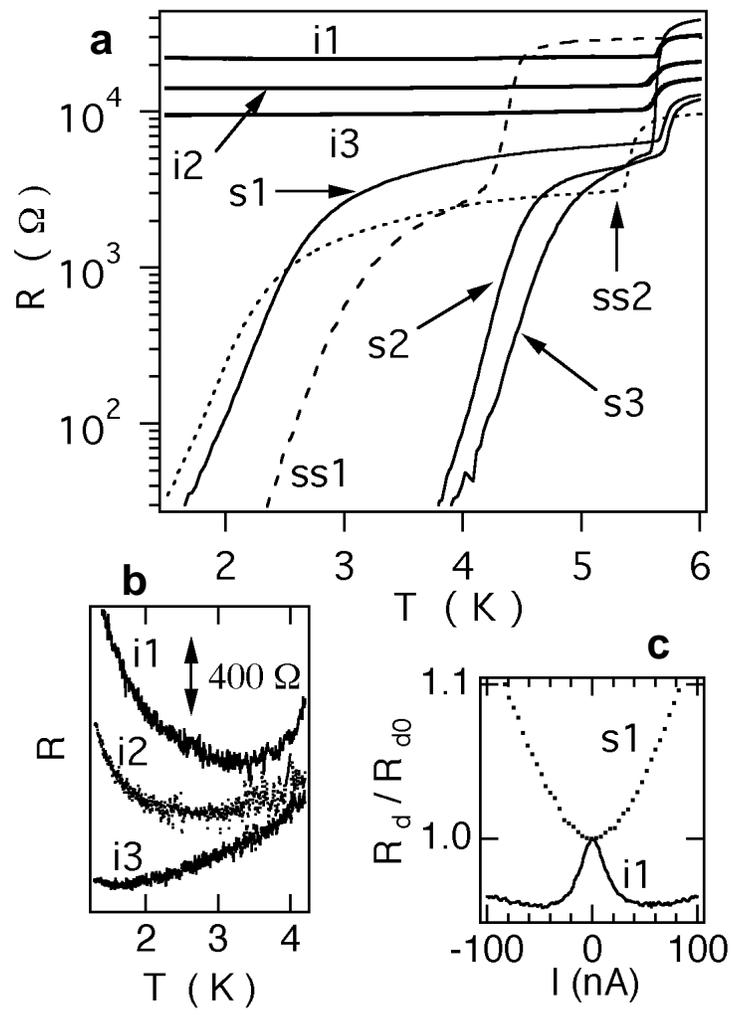

Fig.3